\documentclass[prd,nofootinbib,showpacs,preprint]{revtex4}
\usepackage[T1]{fontenc}
\usepackage{amsmath,amssymb}
\usepackage{epsfig}
\usepackage{graphicx}
\usepackage[usenames,dvipsnames]{color}
\usepackage{slashed}
\usepackage[colorlinks,citecolor=blue]{hyperref}
\usepackage{pdfpages}
\usepackage{color}

\begin{document}

\title{Common Origin of 3.55 keV X-Ray Line and Galactic Center Gamma Ray Excess in a Radiative Neutrino Mass Model}
\author{Debasish Borah}
\email{dborah@tezu.ernet.in}
\affiliation{Department of Physics, Tezpur University, Tezpur - 784028, India}
\author{Arnab Dasgupta}
\email{arnabdasgupta28@gmail.com}
\affiliation{Centre for Theoretical Physics, Jamia Millia Islamia - Central University, Jamia Nagar, New Delhi - 110025, India}
\author{Rathin Adhikari}
\email{rathin@ctp-jamia.res.in}
\affiliation{Centre for Theoretical Physics, Jamia Millia Islamia - Central University, Jamia Nagar, New Delhi - 110025, India}

\begin{abstract}

We attempt to simultaneously explain the recently observed 3.55 keV X-ray line in the analysis of XMM-Newton telescope data and the galactic center gamma ray excess observed by the Fermi gamma ray space telescope within an abelian gauge extension of standard model. We consider a two component dark matter scenario with tree level mass difference 3.55 keV such that the heavier one can decay into the lighter one and a photon with energy 3.55 keV. The lighter dark matter candidate is protected from decaying into the standard model particles by a remnant $Z_2$ symmetry into which the abelian gauge symmetry gets spontaneously broken. If the mass of the dark matter particle is chosen to be within $31-40$ GeV, then this model can also explain the galactic center gamma ray excess if the dark matter annihilation into $b\bar{b}$ pairs has a cross section of $\langle \sigma v \rangle \simeq (1.4-2.0) \times 10^{-26} \; \text{cm}^3/\text{s}$. We constrain the model from the requirement of producing correct dark matter relic density, 3.55 keV X-ray line flux and galactic center gamma ray excess. We also impose the bounds coming from dark matter direct detection experiments as well as collider limits on additional gauge boson mass and gauge coupling. We also briefly discuss how this model can give rise to sub-eV neutrino masses at tree level as well as one-loop level while keeping the dark matter mass at few tens of GeV. We also constrain the model parameters from the requirement of keeping the one-loop mass difference between two dark matter particles below a keV. We find that the constraints from light neutrino mass and keV mass splitting between two dark matter components show more preference for opposite $CP$ eigenvalues of the two fermion singlet dark matter candidates in the model
\end{abstract}

\maketitle

\section{Introduction}
\label{sec:intro}
Recent analysis \cite{Xray1, Xray2} of the observations made by XMM-Newton X-ray telescope have pointed towards a monochromatic X-ray line with approximate energy 3.55 keV in the spectrum of 73 galaxy clusters (For a review of dark matter, please see \cite{Jungman:1995df}). The same line also appears in the \textit{Chandra} observations of the Perseus cluster \cite{Xray1}. In the absence of any astrophysical interpretation of the line due to some atomic transitions, the origin of this X-ray line can be explained naturally by sterile neutrino dark matter with mass approximately 7.1 keV decaying into a photon and a standard model (SM) neutrino. This was pointed out by the authors in \cite{Xray1, Xray2} and subsequently studied within the framework of specific models \cite{Xraysterile}. Several other particle physics explanations of the X-ray line have also been put forward in \cite{Xrayothers1, Xrayothers2}. Although most of the particle physics explanations consider late time decay or annihilation of multi-keV dark matter particles as the origin of the X-ray line, there have also been a few discussions on the scenario where the X-ray line can be generated by transitions between electroweak scale dark matter states with keV mass splittings \cite{Xrayweakscale, Falkowski:2014sma}. In spite of the fact that both keV scale as well as weak scale dark matter candidates can explain the same signal, their implications in cosmology and astrophysical structure formation can be very different. As stated in \cite{Xray1, Xray2}, a keV scale sterile neutrino should have mixing with the SM neutrinos of the order $\approx 10^{-11}-10^{-10}$ to explain the X-ray line. Such a tiny mixing prevents the sterile neutrinos from entering thermal equilibrium in the early Universe, making it necessary to have some additional physics responsible for the production of sterile neutrinos. However, electroweak scale dark matter particles can be thermally populated in the early Universe due to their sizable interactions either through gauge bosons, Higgs portals or fermion portals etc. These scenarios are studied in the context of so-called weakly interacting massive particle (WIMP). The fact that the WIMP annihilation cross section turns out to be almost equal to the annihilation cross section of thermal dark matter in order to produce the correct dark matter relic abundance observed by the Planck experiment \cite{Planck13} 
\begin{equation}
\Omega_{\text{DM}} h^2 = 0.1187 \pm 0.0017
\label{dm_relic}
\end{equation}
is known as the \textit{WIMP Miracle}. In the above equation \eqref{dm_relic}, $\Omega$ is the density parameter and $h = \text{(Hubble Parameter)}/100$ is a parameter of order unity.

Motivated by the possibility of explaining the origin of 3.55 keV X-ray line within WIMP dark matter framework \cite{Xrayweakscale, Falkowski:2014sma}, here we consider an abelian gauge extension of SM with two Majorana fermion dark matter candidates with a keV mass splitting. The UV complete model, originally proposed by \cite{Adhikari:2008uc} and later studied in the context of dark matter and eV scale sterile neutrino in \cite{Borah:2012qr} and \cite{Borah:2014} respectively, can naturally explain dark matter and the origin of tiny neutrino masses. Sub-eV scale SM neutrino masses arise both at tree level as well as one-loop level with dark matter particles running inside the loops, a framework more popularly known as "scotogenic" model \cite{Ma:2006km}. Recently, this model was also studied \cite{arnabborah2} in the context of explaining the galactic center gamma ray excess observed by the Fermi Gamma Ray Space Telescope \cite{GCfeb}. The abelian gauge charges of the SM as well as beyond SM fields are chosen in such a way that the model is free from gauge anomalies and the abelian gauge symmetry $U(1)_X$ gets spontaneously broken down to a remnant $Z_2$ symmetry so that the lightest $Z_2$-odd particle is stable and hence can be a dark matter candidate. As studied in details in \cite{Borah:2012qr}, this model has several dark matter candidates namely, fermion singlet, fermion triplet, scalar singlet and scalar doublet. Scalar dark matter phenomenology is similar to the Higgs portal models discussed extensively in the literature. In these scenarios, the scalar dark matter self-annihilates into the Standard Model (SM) particles through the Higgs boson. Co-annihilations through gauge bosons can also play a role if the CP even and CP odd components of the neutral Higgs have a tiny mass difference as discussed recently in \cite{arnabborah} within the context of a different model. 

Instead of pursuing Higgs portal like scalar dark matter scenarios in the model, we study the fermionic dark matter sector. This is also relevant to our discussion on the origin of 3.55 keV X-ray line. This is because, if transition between two semi-degenerate weak scale dark matter candidates with keV mass splitting is the origin of the X-ray line, then the dark matter candidates have to be fermions as one scalar decaying into another scalar and a photon does not conserve spin. As we will see in the next section, our model has two fermion singlet dark matter candidates with different gauge charges and two fermion triplet dark matter candidates with the same gauge charge. We can choose either a triplet-singlet or a singlet-singlet combination of two semi-degenerate dark matter candidates to explain dark matter abundance as well as the origin of $3.55$ keV X-ray line simultaneously. However, the neutral component of fermion triplet needs to be very heavy ($ 2.28-2.42 \; \text{TeV} $) in order to reproduce correct dark matter relic density \cite{Ma:2008cu}. To allow the possibility of low mass dark matter, we therefore confine our discussion to fermion singlet dark matter in this work. That is, we explore the possibility of two fermion singlet dark matter candidates with keV mass splitting in this model which can simultaneously give rise to the 3.55 keV X-ray line and satisfy experimental bounds on dark matter relic density as well as direct detection cross section. Such fermion singlet dark matter particle will self-annihilate through the abelian vector boson $X$ into SM particles. We also incorporate the collider constraints on such additional vector boson and its gauge coupling. We find that, although the relic density and direct detection constraints allow a significant region of the parameter space, the collider constraints reduce the parameter space into the s-wave resonance region where the gauge boson mass is approximately twice that of dark matter mass. We constrain the model further by incorporating the bound from X-ray line data on the decay width of the heavier dark matter particle. We then check whether the same model can also give rise to the galactic center (GC) gamma ray excess observed by the Fermi Gamma Ray Space Telescope which has a feature similar to annihilating dark matter \cite{GCfeb}. Finally, we briefly discuss whether the chosen dark matter masses are compatible with sub-eV SM neutrino masses and also constrain the model parameters in order to keep the one-loop mass splitting between two dark matter particles below keV scale.

This paper is organized as follows: in section \ref{model}, we briefly discuss the model. In section \ref{sec:darkmatter}, we discuss two component dark matter scenario as a source of 3.55 keV X-Ray line and GC gamma ray excess taking into account all necessary experimental constraints. In section \ref{neutrino}, we discuss the compatibility of light singlet fermion dark matter with neutrino mass and in section \ref{split}, we discuss the one-loop mass splitting between two dark matter candidates. Finally, we conclude in section \ref{conclude}. 

\section{The Model}
\label{model}
The model which we take as a starting point of our discussion was first proposed in \cite{Adhikari:2008uc} which has fermion content shown in table \ref{table1}. The scalar content of the model is modified in our present work with a different $U(1)_X$ gauge charge for the singlet scalar $\chi_2$ and a newly added scalar singlet $\chi_5$ as shown in table \ref{table2}.

\begin{table}
\centering
\begin{tabular}{|c|c|c|c|}
\hline
Particle & $SU(3)_c \times SU(2)_L \times U(1)_Y$ & $U(1)_X$ & $Z_2$ \\
\hline
$ (u,d)_L $ & $(3,2,\frac{1}{6})$ & $n_1$ & + \\
$ u_R $ & $(\bar{3},1,\frac{2}{3})$ & $\frac{1}{4}(7 n_1 -3 n_4)$ & + \\
$ d_R $ & $(\bar{3},1,-\frac{1}{3})$ & $\frac{1}{4} (n_1 +3 n_4)$ & +\\
$ (\nu, e)_L $ & $(1,2,-\frac{1}{2})$ & $n_4$ & + \\
$e_R$ & $(1,1,-1)$ & $\frac{1}{4} (-9 n_1 +5 n_4)$ & + \\
\hline
$N_R$ & $(1,1,0)$ & $\frac{3}{8}(3n_1+n_4)$ & - \\
$\Sigma_{1R,2R} $ & $(1,3,0)$ & $\frac{3}{8}(3n_1+n_4)$ & - \\
$ S_{1R}$ & $(1,1,0)$ & $\frac{1}{4}(3n_1+n_4)$ & + \\
$ S_{2R}$ & $(1,1,0)$ & $-\frac{5}{8}(3n_1+n_4)$ & - \\
\hline
\end{tabular}
\label{table1}
\caption{Fermion Content of the Model}
\end{table}

The third column in table \ref{table1} shows the $U(1)_X$ gauge charges of various fields which satisfy the gauge anomaly matching conditions. The charges of the scalar fields in table \ref{table2} are chosen according to the desired neutrino and dark matter phenomenology. The Higgs content chosen in the model is not arbitrary and is needed, which leads to the possibility of radiative neutrino masses as well as a remnant $Z_2$ symmetry in a manner proposed in \cite{Ma:2006km}. In this model, the quarks couple to scalar $\Phi_1$ and charged leptons to the scalar $\Phi_2$ whereas $(\nu, e)_L$ couples to $N_R, \Sigma_R$ through $\Phi_3$ and to $S_{1R}$ through $\Phi_1$. The Lagrangian which can be constructed from the above particle content has an automatic $Z_2$ symmetry and hence the model provides a stable cold dark matter candidate in terms of the lightest odd particle under this $Z_2$ symmetry. The $Z_2$ transformations of the fields are shown in the fourth column of table \ref{table1} and table \ref{table2}.
\begin{table}
\centering
\begin{tabular}{|c|c|c|c|}
\hline
Particle & $SU(3)_c \times SU(2)_L \times U(1)_Y$ & $U(1)_X$ & $Z_2$ \\
\hline
$ (\phi^+,\phi^0)_1 $ & $(1,2,-\frac{1}{2})$ & $\frac{3}{4}(n_1-n_4)$ & + \\
$ (\phi^+,\phi^0)_2 $ & $(1,2,-\frac{1}{2})$& $\frac{1}{4}(9n_1-n_4)$ & + \\
$(\phi^+,\phi^0)_3 $ & $(1,2,-\frac{1}{2})$& $\frac{1}{8}(9n_1-5n_4)$ & - \\
\hline
$ \chi_1 $ & $(1,1,0)$ & $-\frac{1}{2}(3n_1+n_4)$ & + \\
$ \chi_2 $ & $(1,1,0)$ & $-\frac{5}{4}(3n_1+n_4)$ & + \\
$ \chi_3 $ & $(1,1,0)$ & $-\frac{3}{8}(3n_1+n_4)$ & - \\
$ \chi_4 $ & $(1,1,0)$ & $-\frac{3}{4}(3n_1+n_4)$ & + \\
$ \chi_5 $ & $(1,1,0)$ & $(3n_1-n_4)$ & + \\
\hline
\end{tabular}
\label{table2}
\caption{Scalar Content of the Model}
\end{table}
The scalar Lagrangian relevant for future discussion can be written as 
\begin{equation}
V_s \supset f_3 \chi_1 \chi^{\dagger}_3\Phi^{\dagger}_1\Phi_3 + f_5 \chi^{\dagger}_3\chi_4 \Phi^{\dagger}_3 \Phi_2 +f_6 (\Phi^{\dagger}_1 \Phi^{\dagger}_3) \chi_3 \chi_5
\label{scalarnu}
\end{equation}
Similarly, the relevant part of the Yukawa Lagrangian for the model can be written as 
$$ \mathcal{L}_Y \supset y \bar{L} \Phi^{\dagger}_1 S_{1R} + h_N \bar{L} \Phi^{\dagger}_3 N_R + h_{\Sigma}  \bar{L}\Phi^{\dagger}_3 \Sigma_R + f_N N_R N_R \chi_4+ f_S S_{1R} S_{1R} \chi_1 $$
\begin{equation}
+ f_{\Sigma} \Sigma_R \Sigma_R \chi_4 + f_{S2} S_{2R} S_{2R} \chi^{\dagger}_2 + f_{12} S_{1R} S_{2R} \chi^{\dagger}_3
\label{yukawa} 
\end{equation}
Let us denote the vacuum expectation values (vev) of various neutral scalar fields as $ \langle \phi^0_{1,2} \rangle = v_{1,2}, \; \langle \chi^0_{1,2,4,5} \rangle  =u_{1,2,4,5}$. The additional $U(1)_X$ gauge boson mass, which is relevant for dark matter phenomenology is given by 
\begin{equation}
 M^2_X = 2g^2_X (-\frac{3M^2_W}{8g^2_2}(9n_1-n_4)(n_1-n_4)+\frac{1}{16}(3n_1+n_4)^2(4u^2_1+25u^2_2+9u^2_4)+(3n_1-n_4)^2 u^2_5)
\label{mxmass}
\end{equation}
where $g_X$ is the $U(1)_X$ gauge coupling. For simplicity, the mixing between the neutral electroweak gauge bosons and the additional $U(1)_X$ gauge boson is chosen to be zero which gives rise to the following constraint
\begin{equation}
3(n_4-n_1)v^2_1 = (9n_1-n_4)v^2_2
\label{zeromixeq}
\end{equation}
which further implies $1 < n_4/n_1 <9 $.

\section{Singlet Fermion Dark Matter}
\label{sec:darkmatter}
The relic abundance of a dark matter particle $\chi$ is given by the Boltzmann equation
\begin{equation}
\frac{dn_{\chi}}{dt}+3Hn_{\chi} = -\langle \sigma v \rangle (n^2_{\chi} -(n^{\text{eq}}_{\chi})^2)
\end{equation}
where $n_{\chi}$ is the number density of the dark matter particle $\chi$ and $n^{\text{eq}}_{\chi}$ is the number density when $\chi$ was in thermal equilibrium. $H$ is the Hubble expansion rate of the Universe and $ \langle \sigma v \rangle $ is the thermally averaged annihilation cross section of the dark matter particle $\chi$. In terms of partial wave expansion $ \langle \sigma v \rangle = a +b v^2$. Numerical solution of the Boltzmann equation above gives \cite{Kolb:1990vq}
\begin{equation}
\Omega_{\chi} h^2 \approx \frac{1.04 \times 10^9 x_F}{M_{Pl} \sqrt{g_*} (a+3b/x_f)}
\label{eq:relic}
\end{equation}
where $x_f = m_{\chi}/T_f$, $T_f$ is the freeze-out temperature, $g_*$ is the number of relativistic degrees of freedom at the time of freeze-out. Dark matter particles with electroweak scale mass and couplings freeze out at temperatures approximately in the range corresponding to $x_f \approx 20-30$. More generally, $x_f$ can be calculated from the relation 
\begin{equation}
x_f = \ln \frac{0.038gM_{\text{Pl}}m_{\chi}<\sigma v>}{g_*^{1/2}x_f^{1/2}}
\label{xf}
\end{equation}
where $g$ is the number of internal degrees of freedom of the dark matter particle $\chi$ and $M_{\text{Pl}}$ is the Planck mass. The thermal averaged annihilation cross section $\langle \sigma v \rangle$ is given by \cite{Gondolo:1990dk}
\begin{equation}
\langle \sigma v \rangle = \frac{1}{8m^4_{\chi}T K^2_2(m_{\chi}/T)} \int^{\infty}_{4m^2_{\chi}}\sigma (s-4m^2_{\chi})\surd{s}K_1(\surd{s}/T) ds
\label{eq:sigmav}
\end{equation}
where $K_i$'s are modified Bessel functions of order $i$, $m_{\chi}$ is the mass of Dark Matter particle and $T$ is the temperature.

There are two singlet fermions $N_R, S_{2R}$ in this model which are odd under the remnant $Z_2$ symmetry and hence can be a dark matter candidate, if lightest among all the $Z_2$-odd particles. We consider a scenario where $S_{2R}$ is the lightest and $N_R$ is the next to lightest $Z_2$-odd particle of the model. If the lifetime of $N_R$ is very high, longer than the present age of the Universe, then both $S_{2R}$ and $N_R$ can contribute to the present abundance of dark matter. From the field content and their gauge charges, one can see that there is no term in the Lagrangian which involves both $S_{2R}$ and $N_R$. Also there is no scalar which couples to both $S_{2R}$ and $N_R$. Thus, there is no co-annihilating processes between $S_{2R}$ and $N_R$ which can contribute to the dark matter relic abundance. Hence, one can calculate the relic abundance of $S_{2R}$ and $N_R$ separately, keeping them decoupled. To calculate the relic density of either $S_{2R}$ or $N_R$, we need to find out its annihilation cross-section to standard model particles. For zero $Z-X$ mixing, the dominant annihilation channel is the one with $X$ boson mediation. Since the singlet fermions are of Majorana type, they have only axial coupling to the vector boson. The annihilation cross-section of Majorana singlet fermion into SM fermion anti-fermion pairs $f\bar{f}$ through s-channel $X$ boson \cite{Berlin:2014tja} can be written as
\begin{align}
\sigma &= \frac{n_c}{12\pi s\left[(s-m^2_{X})^2+M^2_X\Gamma^2_X\right]}\bigg{[}\frac{1-4m^2_f/s}{1-4M^2_X/s}\bigg{]}^{1/2}\times \nonumber \\
&\bigg{[}g^2_{fa}g^2_{\chi a}\bigg{(}4m^2_{\chi}\bigg{[}m^2_f\bigg{(}7-\frac{6s}{M^2_X}+\frac{3s^2}{M^4_X}\bigg{)}-s\bigg{]}+s(s-4m^2_f)\bigg{)} \nonumber \\
&+g^2_{fv}g^2_{\chi a}(s+2m^2_f)(s-4m^2_{\chi})\bigg{]}
\end{align}
Expanding in powers of $v^2$ gives 
$\sigma v $ in the form $a + b v^2$ where $a$ and $b$ are given by
\begin{align}
a &= \frac{n_c g^2_{fa}m^2_fg^2_{\chi a}m^2_{\chi}}{24\pi^2m^2_{\chi}((M^2_X - 4 m^2_{\chi})^2+M^2_X\Gamma^2_X)}\sqrt{1-\frac{m^2_f}{m^2_{\chi}}}\bigg{(}-36 + 48 \frac{m^2_{\chi}}{m^2_f}-96\frac{m^2_{\chi}}{M^2_X}+192\frac{m^4_{\chi}}{M^4_X}\bigg{)} \nonumber \\
b &= a\bigg{[}-\frac{1}{4}+\frac{2m^2_{\chi}(M^2_X-4m^2_\chi)}{(M^2_X-4m^2_\chi)^2+M^2_X\Gamma^2_X}+\frac{1}{8(m^2_\chi-m^2_f)m^2_f}\nonumber \\
&+\frac{\left(-16+2\frac{g^2_{fv}}{g^2_{fa}}+28\frac{m^2_\chi}{m^2_f}+4\frac{g^2_{fv}m^2_\chi}{g^2_{fa}m^2_f}-24\frac{m^2_\chi}{M^2_X}+96\frac{m^4_\chi}{M^4_X}\right)}{\left(-36 + 48 \frac{m^2_{\chi}}{m^2_f}-96\frac{m^2_{\chi}}{M^2_X}+192\frac{m^4_{\chi}}{M^4_X}\right)}\bigg{]}
\end{align}

The Decay width of the $X$ boson denoted by $\Gamma_X$ is given by
\begin{align}
\Gamma_{X\rightarrow \chi\overline{\chi}} &= \frac{n_cM_Xg^2_X}{12\pi S}\bigg{[}1-\frac{4m^2_{\chi}}{m^2_{X}}\bigg{]}^{3/2} \nonumber \\
\Gamma_{X \rightarrow f \overline{f}} &= \sum_{f} \frac{n_cM_X}{12\pi S}\bigg{[}1-\frac{4m^2_f}{M^2_X}\bigg{]}^{1/2}\bigg{[}g^2_{fa}\bigg{(}1-\frac{4m^2_f}{m^2_{X}}\bigg{)} \nonumber \\ 
&+g^2_{fv}\bigg{(}1+2\frac{m^2_f}{M^2_X}\bigg{)}\bigg{]}
\end{align}
The mass of the gauge boson $X$ in the above expressions is given by equation (\ref{mxmass}). For simplicity, we assume $u_1 = u_2 = u_4 = u$ such that the mass of $X$ boson can be written as
\begin{align}
M^2_X &= 2g^2_X\bigg{[}-3\frac{m^2_W}{8g^2_2}(9n_1-n_4)(n_1-n_4)+\frac{19}{8}(3n_1+n_4)^2u^2 +(3n_1-n_4)^2 u^2\bigg{]}
\label{eq:mX}
\end{align}
The couplings $g_{fv}, g_{fa}, g_{\chi v}, g_{\chi a}$ of fermions and dark matter to $X$ boson are tabulated in the table \ref{table:coupling}.
\begin{table}[!h]
\label{table:coupling}
\centering
\begin{tabular}{|c|c|c|c|}
\hline
& $n_c$ & $g_{fv}/g_X$ & $g_{fa}/g_X$ \\
\hline \hline
$l=e,\mu,\tau$ & 1 & $\frac{9}{8}\left(n_4-n_1\right)$ & $\frac{1}{8}\left(n_4-9n_1\right)$   \\
$\nu_l$ & 1 & $\frac{n_4}{2}$ & $-\frac{n_4}{2}$ \\
$U=u,c$ & 3 & $\frac{1}{8}(11n_1-n_4)$& $\frac{3}{8}(n_1-n_4)$ \\
$D=d,s,b$ & 3 & $\frac{1}{8}(5n_1+3n_4)$ & $\frac{3}{9}(n_4-n_1)$ \\
$N_R$ & 1 & 0 & $\frac{3}{8}(3n_1+n_4)$ \\
$S_{2R}$ & 1 & 0 & $-\frac{5}{8}(3n_1+n_4) $ \\
\hline
\end{tabular}
\caption{Couplings of SM particles and dark matter to the vector boson $X$}
\end{table}

\begin{figure*}[!h] 
\centering
\begin{tabular}{c}
\epsfig{file=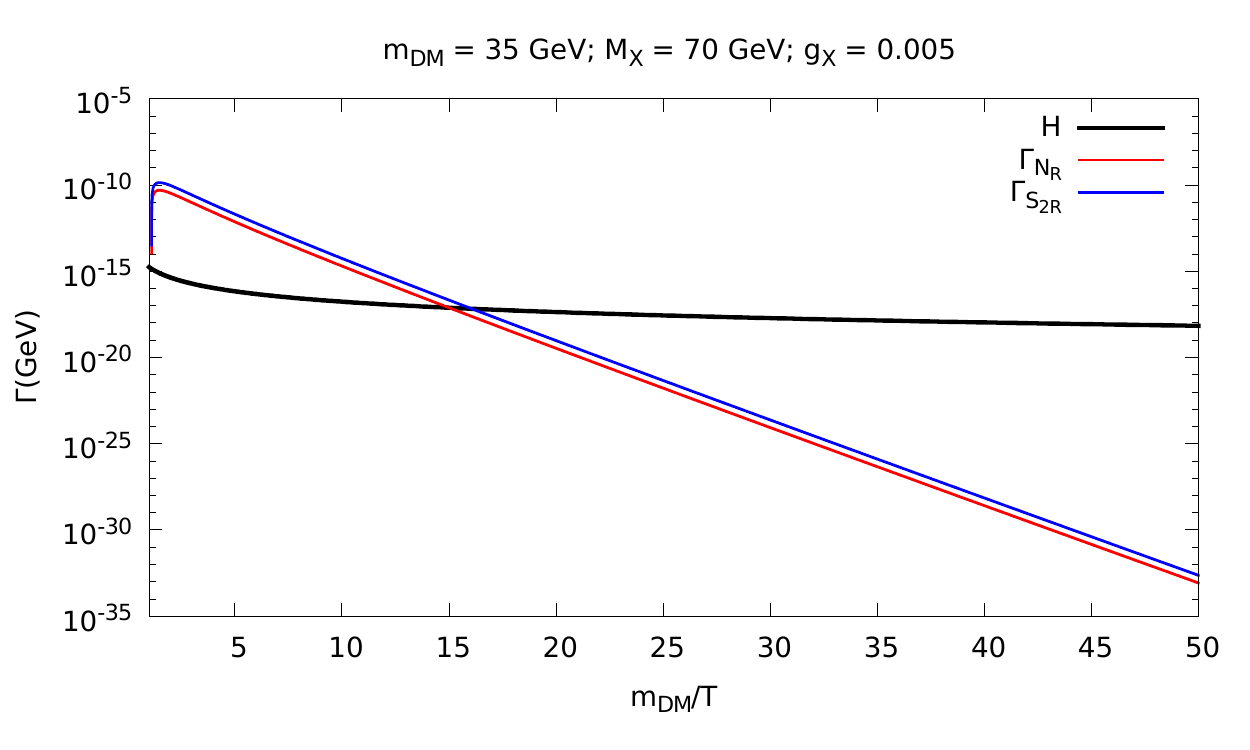,width=1.0\linewidth,clip=}\\
\end{tabular}
\caption{Comparison of self-interaction rates of dark matter candidates $(N_R, S_{2R})$ with $m_{S_{2R}} = m_{\text{DM}} = 35$ GeV and $m_{N_R}-m_{S_{2R}} = 3.55$ keV with the Hubble expansion rate $H$ for $M_X = 70$ GeV, $g_X = 0.005$.}
\label{fig1}
\end{figure*}

\begin{figure*}[!h] 
\centering
\begin{tabular}{c}
\epsfig{file=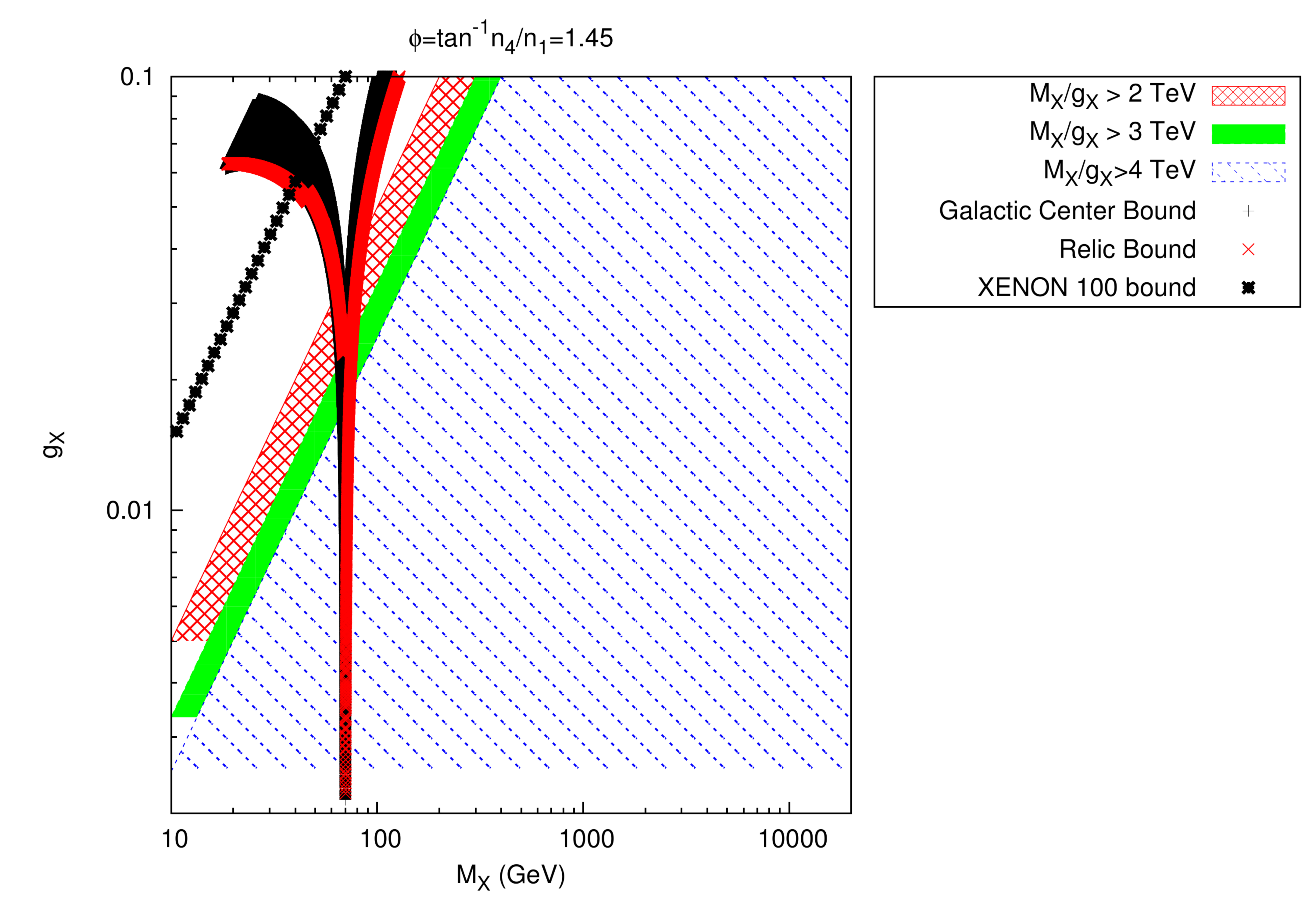,width=1.0\linewidth,clip=}\\
\end{tabular}
\caption{Parameter space in the $g_X-M_X$ plane for two-component dark matter $(N_R, S_{2R})$ scenario with $m_{S_{2R}} = 35$ GeV and $m_{N_R}-m_{S_{2R}} = 3.55$ keV. The red-hatched, green and blue dot-dashed regions correspond to the allowed region after the constraints on $M_X/g_X$ are imposed. The area to the left of the black line is ruled out by XENON100 bounds on direct detection cross section. The solid red region corresponds to the parameter space favored by the relic density constraint. The solid black region corresponds to the parameter space favoured by galactic centre gamma ray excess.}
\label{fig2}
\end{figure*}

\begin{figure*}[!h] 
\centering
\begin{tabular}{c}
\epsfig{file=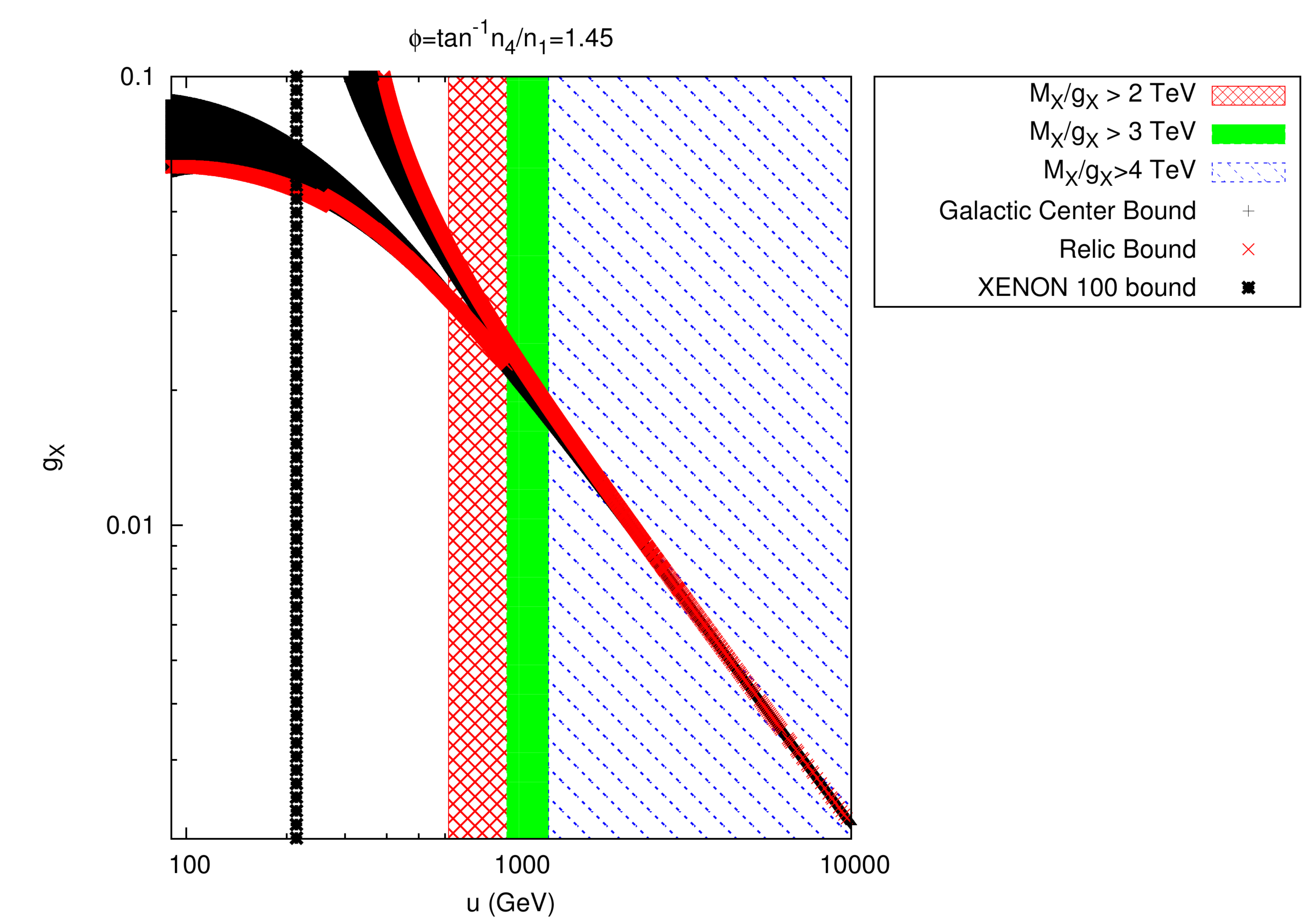,width=1.0\linewidth,clip=}\\
\end{tabular}
\caption{Parameter space in the $g_X-u$ plane for two-component dark matter $(N_R, S_{2R})$ scenario with $m_{S_{2R}} = 35$ GeV and $m_{N_R}-m_{S_{2R}} = 3.55$ keV. The red-hatched, green and blue dot-dashed regions correspond to the allowed region after the constraints on $M_X/g_X$ are imposed. The area to the left of the black line is ruled out by XENON100 bounds on direct detection cross section. The solid red region corresponds to the parameter space favored by the relic density constraint. The solid black region corresponds to the parameter space favoured by galactic centre gamma ray excess.}
\label{fig3}
\end{figure*}

\begin{figure*}[!h] 
\centering
\begin{tabular}{c}
\epsfig{file=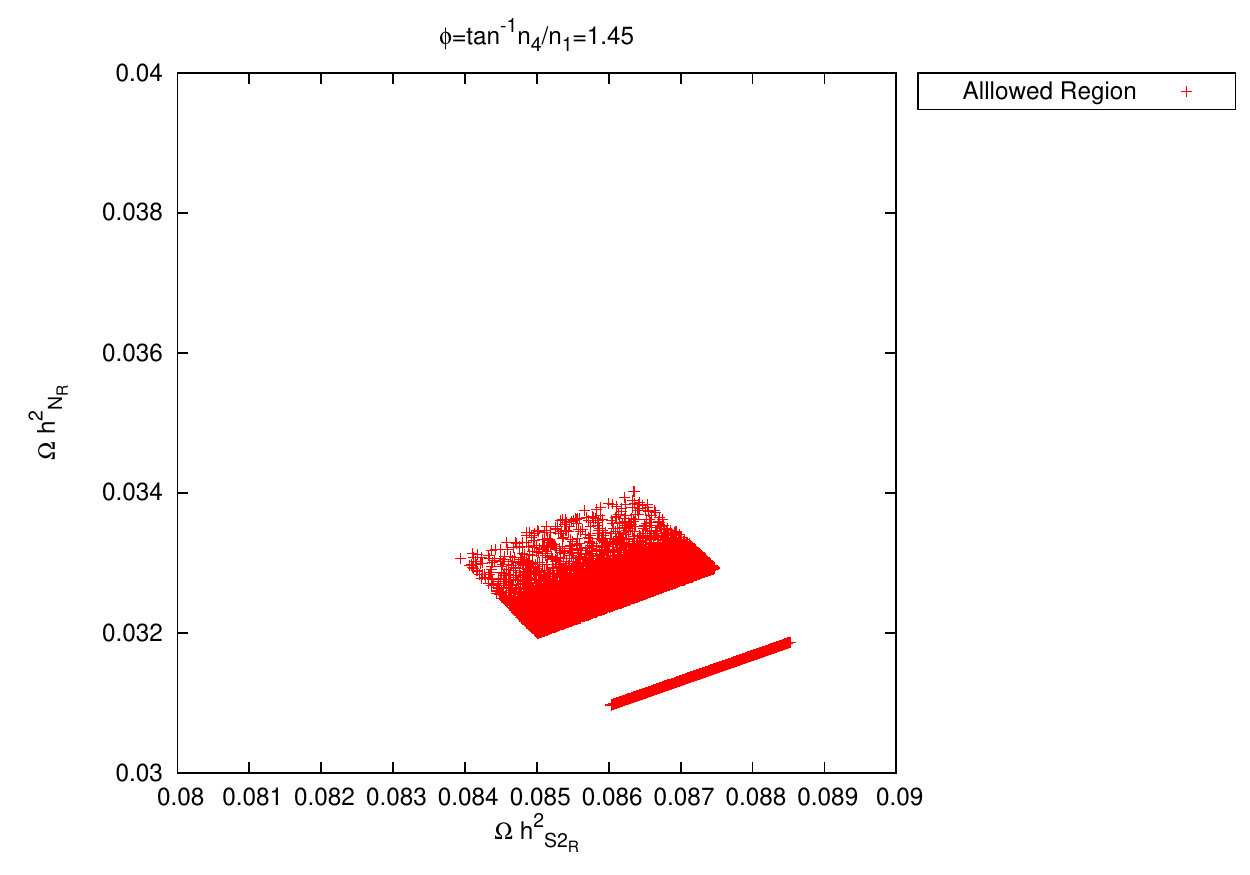,width=1.0\linewidth,clip=}\\
\end{tabular}
\caption{Relative contribution of $N_R$ and $S_{2R}$ to dark matter relic abundance for $m_{S_{2R}} = 35$ GeV and $m_{N_R}-m_{S_{2R}} = 3.55$ keV.}
\label{fig4}
\end{figure*}

Using the couplings given in table \ref{table:coupling}, we first check whether there exists a freeze-out temperature $T_f$ for singlet fermionic dark matter such that for $T > T_f$, the singlet fermions enter thermal equilibrium. This is crucial in order to use the standard relic abundance formula given by equation (\ref{eq:relic}) for WIMP dark matter. For this we need to calculate the annihilation rate of singlet dark matter particles and compare with the Hubble expansion rate of the Universe. To calculate the interaction rate and hence annihilation cross section in our case, we fix the gauge charges $n_1, n_4$ but vary the $U(1)_X$ gauge coupling $g_X$ and gauge boson mass $M_X$. Similar to our approach in \cite{Borah:2012qr, arnabborah2}, here also we choose a specific value of $ n_1$ from which $n_4$ can be determined using the normalization relation $n^2_1 +n^2_4 = 1$. Using the same normalization, the $90\%$ confidence level exclusion on $M_X/g_X$ was shown in \cite{Adhikari:2008uc} where the lowest allowed value of $M_X/g_X$ was found to be approximately $2$ TeV for $\phi = \tan^{-1} (n_4/n_1) = 1.5$. Using this and the normalization relation involving $n_1, n_4$ we determine both $n_1, n_4$. Since $U(1)_X$ gauge charges of all the fields are written in terms of $n_1, n_4$ it is sufficient to choose just these two values to determine all the gauge charges. After fixing dark matter mass as well as $n_{1,4}$, we vary $g_X$ and $u$ and compute the annihilation cross section of dark matter particles. The freeze-out temperature $T_f$ is then calculated numerically by using the equation 
\begin{align}
e^{x_f} - \ln \frac{0.038gm_{PL}m_{\chi}<\sigma v>}{g_*^{1/2}x_f^{1/2}} &= 0
\end{align}
This is a simplified form of equation (\ref{xf}). For a fixed value of dark matter mass $m_{\chi}$, the annihilation cross section $\sigma$ depends upon $g_X, M_X$. For a particular pair of $g_X$ and $M_X$, we use this value of $x_f$ and compute the relic abundance using equation (\ref{eq:relic}). 

To show that the dark matter candidates in our model have gone through this generic freeze-out process, we compare their interaction rates with the Hubble expansion rate of the Universe. The interaction rate is given by $\Gamma = n \langle \sigma v \rangle$, where $n$ is the number density, and $\langle \sigma v \rangle$ can be calculated using equation \eqref{eq:sigmav}. For a non-relativistic dark matter particle of mass $m$, the equilibrium number density is given by 
\begin{equation}
n = g \left ( \frac{mT}{2\pi} \right )^{3/2} e^{-m/T}
\label{eq:neqb}
\end{equation}
where we have taken the chemical potential to be zero. Here, $g=2$ for Majorana fermion dark matter. On the other hand, the Hubble expansion parameter for the early radiation dominated epoch can be written as 
\begin{equation}
H = 1.66 g^{1/2}_* \frac{T^2}{M_{\text{Pl}}} 
\label{eq:hubble}
\end{equation}
We plot $\Gamma = n \langle \sigma v \rangle$ as well as $H$ as a function of $m/T$ for two dark matter particles with masses $m_{S_{2R}} = m_{\text{DM}} = 35$ GeV and $m_{N_R}-m_{S_{2R}} = 3.55$ keV for gauge boson mass $M_X = 70$ GeV and gauge coupling $g_X = 0.005$. This is shown in figure \ref{fig1}. The point at which the interaction rate $\Gamma$ falls below the Hubble expansion rate $H$ corresponds to the freeze-out temperature $T_f$. From figure \ref{fig1}, we see that this crossover occurs at $m_{DM}/T \approx 15$, which corresponds to freeze-out temperature $T_f \approx 2.33$ GeV. Calculation of dark matter self-annihilation cross section also allows us to find out the parameter space giving rise to the correct relic abundance. This parameter space in $g_X-M_X$ and $g_X-u$ planes are shown in \ref{fig2} and \ref{fig3} respectively.

We then consider the bound on dark matter nucleon scattering cross section from direct detection experiments. Since both the dark matter candidates in our model are Majorana fermions, the vector current vanishes and they give rise to spin dependent scattering cross section with nuclei. However, spin independent scattering can arise if there exist scalar mediated interactions between dark matter and nucleons. In the limit where the mixing between singlet scalars $\chi_2, \chi_4$ and the scalar doublet $\Phi_1$ is negligible, we can consider only the gauge boson mediated spin dependent scattering between dark matter and nucleons. The latest upper bound on this scattering cross section comes from the XENON100 experiment \cite{Aprile:2013doa}. The expression for this spin dependent scattering of dark matter particles off nuclei through t-channel mediation of X boson can be written as
\begin{align}
\sigma_{SD} &= \frac{4\mu^2_{\chi N}}{\pi M^4_X}g^4_{\chi a}J_N(J_N+1)
\bigg{(}\frac{\langle S_p\rangle}{J_N}(2\Delta^{(p)}_u +\Delta^{(p)}_d) \nonumber \\ 
&+\frac{\langle S_n\rangle}{J_N}(2\Delta^{(n)}_d+\Delta^{(n)}_u)\bigg{)}^2
\end{align}
where 
$$ \mu_{\chi N} = \frac{m_\chi m_N}{m^2_{\chi}+m^2_N}$$ 
and $J_N$ is the spin of the Xenon nucleus.
The standard values of the nuclear quark content are$\Delta^{(p)}_u=\Delta^{(n)}_d=0.84$ and $\Delta^{(n)}_u=\Delta^{(p)}_d=-0.43$ \cite{pdg}.
The average spins $\langle S_p\rangle$ and $\langle S_n \rangle$ of the Xenon nucleus are taken from \cite{Aprile:2013doa} as given in table \ref{table1:nuc}.
\begin{table}[!h]
\centering
\begin{tabular}{|c|c|c|}
\hline \hline 
Nucleus & $\langle S_n\rangle$ & $\langle S_p \rangle$ \\
\hline
$^{129}Xe$ & 0.329 & 0.010 \\
$^{131}Xe$ & -0.272 & -0.009 \\
\hline
\end{tabular}
\caption{Average Spin of Xenon Nucleus}
\label{table1:nuc}
\end{table}
The lowest upper bound at $90\%$ confidence level from XENON100 experiment on spin dependent dark matter nuclei cross section is $3.5 \times 10^{-40} \; \text{cm}^2$ for dark matter mass of 45 GeV. Here we take this as a conservative upper bound on direct detection cross section and show the region of parameter space in both $g_X-M_X$ and $g_X-u$ planes which gives rise to this cross section. This gives rise to a solid exclusion line in the figure \ref{fig2} and \ref{fig3} so that the region of parameter space above or towards left of this line is ruled out. Similar to the discussion in our earlier work \cite{arnabborah2}, we also incorporate the collider bounds on $M_X$ and $g_X$. Collider constraints on additional gauge bosons masses with generic SM like gauge couplings force them to be heavier than approximately 2.5 TeV. However, as discussed in \cite{collider}, the bounds on the mass of additional boson $X$ can be relaxed if it has non-negligible coupling to the dark sector. The authors showed that for $X$ decaying into SM particles with branching ratio $90\%$ and $g_X = 0.1$, the lowest allowed value of $M_X/g_X$ is approximately $2.6$ TeV. This limit goes up to 4 TeV and $4.4$ TeV, if $g_X$ is increased to $0.3$ and weak gauge coupling $g$ respectively. To apply these bounds, we calculate the branching ratios of X boson into SM and dark sector particles and find that the maximum branching ratio of X boson into dark matter particles is approximately $8.5\%$. According to the analysis of \cite{collider}, this will correspond to an approximate bound $M_X/g_X > 2.6$ TeV for $g_X = 0.1$. However, these bounds will be weaker if $g_X$ is lowered down into the resonance region that can be seen from figure \ref{fig2} and \ref{fig3}. We apply moderate as well as conservative bounds on $M_X/g_X$ between 2 TeV to 4 TeV and show the portion of parameter space left after that in figure \ref{fig2} and figure \ref{fig3}. 

It can be seen from the figure \ref{fig2} and figure \ref{fig3} that the bounds on $M_X/g_X$ necessarily rules out most of the parameter space in $g_X-M_X$ or $g_X-u$ plane which give correct dark matter properties. Only a narrow region of parameter space near the s-channel resonance $M_X \approx 2m_{DM}$ is left. If dark matter mass is light, a few tens of GeV, then additional bounds from LEP-II experiment will apply on neutral gauge boson and its coupling. The agreement between LEP-II measurements and the standard model predictions forces the mass of additional neutral boson to be greater than 209 GeV or the couplings to be smaller than or of order $10^{-2}$ \cite{pdg}. This will further reduce the parameter space to the region with $g_X \leq 10^{-2}$.

After finding the parameter space which keeps the total abundance of $N_R$ and $S_{2R}$ within the Planck limit on dark matter abundance \eqref{dm_relic}, we also show the relative contribution of $N_R$ and $S_{2R}$ to dark matter relic abundance in figure \ref{fig4}. It can be seen from the figure that $N_R$ can give rise to $26-28\%$ of dark matter relic density whereas $S_{2R}$ gives rise to the rest of it. Since their relative abundances are different, their scattering probability at direct detection experiments will also be different. We have taken that relative factor into account while calculating the dark matter direct detection cross section.
\begin{figure}[h!]
\centering
\epsfig{file=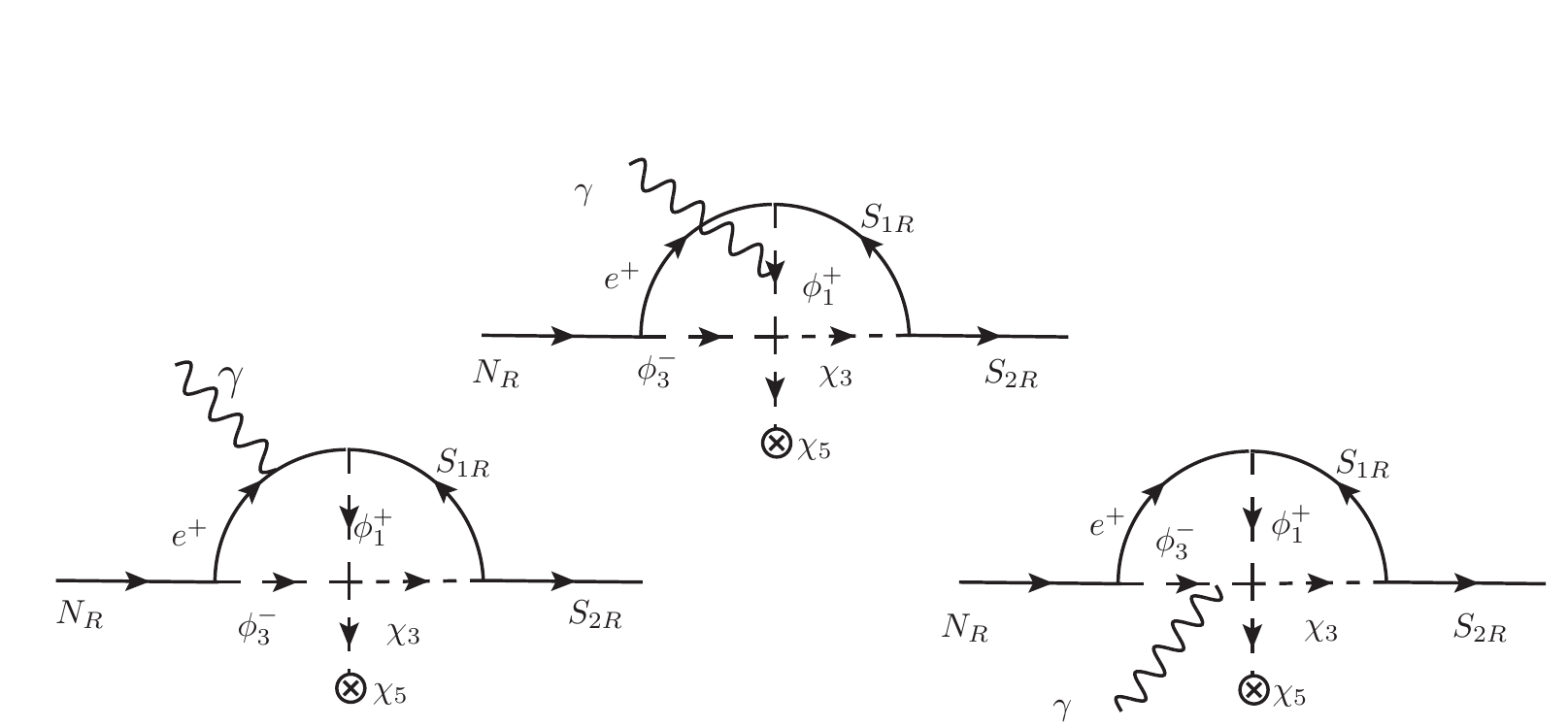,width=0.8\linewidth,clip=}
\caption{Radiative decay of $N_R$ into $S_{2R}$ and a photon $\gamma$.}
\label{fig05}
\end{figure}

In order to fit our model with the observed $3.55$ keV X-Ray line data \cite{Xray1}, we follow the constraint on the 
decay width of the heavier dark matter candidate $N_R$ as obtained in\cite{Falkowski:2014sma}
\begin{equation}
\Gamma (N_R \rightarrow S_{2R} \gamma) \approx 6.2 \times 10^{-47} m_{N_R} \; \text{GeV},
\label{xray}
\end{equation}
where $N_R$ contributes around $50\%$ to dark matter relic abundance. 
Here the dependence on $m_{N_R}$ arises via the number density of dark matter.
In our model, the heavier dark matter particle $N_R$ can decay into the lighter dark matter particle $S_{2R}$ and a photon $\gamma$ only at two-loop level. The corresponding Feynman diagrams can be seen in figure \ref{fig05}. 
In the following we try to make a rough estimate of the decay width due to these two-loop feynmann diagrams, 
assuming all the fields inside the loop to be around TeV scale.  
%
Using our estimate for this decay width expressed in terms of the couplings and also using above constraint \eqref{xray} we obtain 
conservative bound on the product of various couplings as discussed below.

As $N_R$ and $S_{2R}$ are Majorana neutrinos, corresponding to all diagrams in figure \ref{fig05} there will be conjugate diagrams where photon connects to opposite sign particles with respect to diagrams in figure \ref{fig05}. Considering those conjugate diagrams and also considering all heavy masses in the internal line of the diagrams almost degenerate and neglecting electron mass, the decay width can be approximated as

\begin{align}
\Gamma(N_{R}\rightarrow S_{2R} \gamma) &\simeq \left(\frac{m^2_{N_{R}}-m^2_{S_{2R}}}{16\pi m^3_{N_{R}}}\right)
{\left( m^2_{N_{R}}-m^2_{S_{2R}}\right)}^2 \left[F_1^2 + F_2^2\right]
\end{align}
where 
\begin{align}
F_1\simeq 2 \left(\frac{m_{N_{R}}-m_{S_{2R}}}{{m_{\phi^{-}}}^2}\right) I \; ;
F_2\simeq 2 \left(\frac{m_{N_{R}}+m_{S_{2R}}}{{m_{\phi^{-}}}^2}\right) I  \; ;
I \simeq \left(\frac{h_Nf_6 f_{12} \; y\;
u_{5}}{256\pi^4 m_{\phi^{-}}}\right).
\label{eq:decayN}
\end{align}
Here, $F_1$ and $F_2$ correspond to Lorentz invarient form factors connected to electric dipole moment transition and purely 
magnetic moment transition respectively. In general both of them will be contributing to such decays if there is $CP$ violating
 interactions involving $N_R$ as well as $S_{2R}$. However, here we assume that $CP$ is not violated by the interactions involving $N_R$ and 
$S_{2R}$. Then two possible $CP$ eigenvalues $\in\{+i,-i\}$ are possible for Majorana particles like $N_R$ and $S_{2R}$. 
So, either $N_{R}$ and $S_{2R}$ will have same $CP$ eigenvalues or opposite $CP$ eigenvalues. For same $CP$ eigenvalues we get purely electric dipole moment transition for which $F_2=0$  in eq. \eqref{eq:decayN} and for opposite $CP$ eigenvalues we get purely magnetic moment transition \cite{Kayser:1984ge,Pal:1981rm} for which $F_1=0$. Taking the constraint \eqref{xray} into account, we have
\begin{eqnarray}
\Gamma(N_{R}\rightarrow S_{2R} \gamma)  \simeq \left(\frac{{\left(m_{N_{R}}+m_{S_{2R}}\right)}^3}{16\pi m^3_{N_{R}}}\right)
{\Delta k}^3 \left[ F_1^2 + F_2^2\right] \nonumber \\  \sim 6.2 \times 10^{-47} M_{N_R} \; \text{GeV}  
\end{eqnarray}
where
\begin{align}
\Delta k = m_{N_{R}}-m_{S_{2R}}= 3.55 \times 10^{-6} (\textrm{GeV}).
\end{align}
Taking $m_{N_{R}}\approx m_{S_{2R}} \approx 35$ GeV, $m_{\phi^{-}} \approx 500$ GeV and $u_5 \approx 5$ TeV, this constraint can be naturally 
satisfied if
\begin{eqnarray}
h_Nf_6 f_{12}\; y &&\sim 1 \;  (\textrm{for same $CP$ eigenvalues}). \nonumber \\
&&\sim 10^{-14} \; (\textrm{for opposite  $CP$ eigenvalues}).
\label{xrayconstraint}
\end{eqnarray}
As will be discussed in the next sections, to satisfy the constraints from light neutrino masses as well as the one-loop mass splitting 
between $N_R$ and $S_{2R}$, the opposite $CP$ eigenvalues of $N_R$ and $S_{2R}$ seem to be appropriate.

After constraining the model parameters from dark matter, collider as well as the $3.55$ keV X-ray line data, we 
check if the model can explain the galactic center gamma ray excess for the same region of parameter space. Recent analysis \cite{GCfeb} of the Fermi Gamma Ray Space Telescope data has shown an excess of gamma rays with a peak of $1-3$ GeV in the region surrounding the galactic center. 
Also reported by earlier analysis \cite{GCprev}, the spectral shape of the gamma rays has a feature which resembles annihilating dark matter. 
One obtains better fit of the observed gamma ray excess from the annihilation of dark matter to $b\bar{b}$ pairs for dark matter mass $35$ GeV\cite{GCfeb} with the
constraint on the cross-section $\langle \sigma v \rangle = (0.77-3.23)\times 10^{-26}\text{cm}^3/s$\cite{Berlin:2014tja}. Fit with the observed gamma ray excess for 
other masses of the 
dark matter with different annihilation channels has been discussed in reference\cite{GCfeb}. 
In our subsequent analysis we have considered particularly dark matter mass $35$ GeV. 
Since we have two dark matter candidates with a mass difference of $3.55$ keV, we consider the mass of $N_R$ to be $35+3.55 \times 10^{-6}$ GeV and that of $S_{2R}$ to be 35 GeV. We then show in figure \ref{fig2} and \ref{fig3} the region of parameter space in $g_X-M_X$ and $g_X-u$ planes for which the total annihilation cross section of $N_R, S_{2R}$ 
matches the one mentioned above in order to produce the observed gamma ray excess. We include their relative abundance factors while calculating the annihilation cross sections needed to produce galactic center gamma ray excess. The allowed mass of neutral vector boson becomes around 70 GeV which faces severe constraints from LEP-II data and further constrains the coupling 
$g_X \lesssim 10^{-2}$, as discussed earlier. Similarly, we find small allowed parameter space for $g_X\lesssim 10^{-2}$ and
$u \gtrsim 2$ TeV where for simplicity we have assumed $u_1=u_2=u_3=u_4=u$ as mentioned just above equation \eqref{eq:mX}.

It should be noted, taking into account of the recent work \cite{merle1}, that the model of one-loop radiative neutrino mass with dark matter originally proposed by Ma and popularly known as "scotogenic" model \cite{Ma:2006km} suffers from a hierarchy type problem which can spoil the phenomenological success of this model. The problem occurs due to the contributions from heavy Majorana neutrinos through renormalisation group evolution (RGE) to the mass parameters of the scalar fields which are odd under the unbroken $Z_2$ symmetry of the model. If, for some parameter space of the model, the mass parameters of the $Z_2$-odd scalars turn negative at some energy scale, it will break the $Z_2$ symmetry resulting in the loss of a cold dark matter candidate in terms of the lightest $Z_2$-odd particle. Although the model we are studying is an extension of the original Ma's model by a gauge symmetry $U(1)_X$, the effective low energy model below the $U(1)_X$ breaking scale is a Ma type model with heavy Majorana neutrinos and a $Z_2$ symmetry under which several scalar fields are odd. As shown by the authors of \cite{merle1}, one can prevent the mass parameters of the $Z_2$-odd scalars from turning negative if the physical $Z_2$-odd scalar masses are restricted to a certain ranges, typically of the order of the heaviest Majorana neutrino. Since our dark matter candidates are singlet $Z_2$-odd Majorana fermions and instead of $Z_2$-odd scalars, these constraints can be satisfied easily without affecting the dark matter phenomenology discussed above.

\section{Light Neutrino Mass}
\label{neutrino}
The origin of tiny neutrino masses was discussed in details in \cite{Borah:2012qr}. The tiny masses can arise at both tree level as well as one-loop level through the Feynman diagram shown in earlier works \cite{Borah:2012qr, Borah:2014, arnabborah2}. Since out of the three singlet neutrinos $N_R, S_{1R}, S_{2R}$, only $S_{1R}$ gives rise to a Dirac mass term $m_D = y v_1$ for the neutrinos through the vev of $\Phi_1$ (denoted as $v_1$), only one of the neutrinos acquire a non-zero mass at tree level through type I seesaw mechanism \cite{ti}. The tree level mass for the light neutrino in terms of the Dirac mass term and the mass of the heavy singlet neutrino $S_{1R}$ ($M_{S_{1R}} = f_S u_1$) can be written as
\begin{equation}
m_{\nu} \approx \frac{ 2y^2 v_1^2}{f_S u_1}
\label{neutmass}
\end{equation} 
From figure \ref{fig3}, we see that the allowed region from dark matter as well as collider constraints suggest 
$u_1 = u_2 = u_4 = u_5 =u \gtrsim 2$ TeV, so we have taken $u_1 \approx 5$ TeV and $f_su_1 \approx 2$ TeV. Since $v_1 \sim 100$ GeV, for light 
neutrino masses to be at sub-eV scale, the equation (\ref{neutmass}) suggest that the Yukawa couplings $y$ have to be around $3\times10^{-6}$ 
which is approximately same as the electron Yukawa coupling in the SM. The other two SM neutrinos can acquire non-zero masses only when one-loop 
contributions are taken into account. As discussed in \cite{Borah:2012qr}, the  one-loop contribution $(M_\nu)_{ij}$ to 
neutrino mass is given by
\begin{eqnarray} 
({M_\nu)}_{ij} \approx  \frac{f_3 f_5 v_1 v_2 u_1 u_4}{16 \pi^2} \sum_k {h_{N, \Sigma} }_{ik} {h_{N, \Sigma} }_{jk} \left( A_k +{(B_k)}_{ij} \right)
\label{nuradmass}
\end{eqnarray}
Assuming all the scalar masses in the loop diagram to be almost degenerate and written as $m_{sc}$ then 
\begin{eqnarray}
A_k + (B_k)_{ij} \approx m_{2k} \left[\frac{m_{sc}^2 
+ m_{2k}^2 }{m_{sc}^2 \left( m_{sc}^2 - m_{2k}^2 \right)^2 }- \frac{(2-\delta_{ij})\; m_{2k}^2}{\left(m_{sc}^2 - m_{2k}^2 \right)^3}\ln \left( m_{sc}^2/m_{2k}^2 \right)   \right],
\label{scaldeg}
\end{eqnarray}
where $(M_{N, \Sigma})_k = m_{2k}$. For fermion singlet light dark matter, $m_{2k} \ll m_{sc}$ and hence the above expression can be approximated as 
$$ A_k + (B_k)_{ij} \approx \frac{m_{2k}}{m^4_{sc}} $$
The one-loop neutrino mass can be written as
\begin{eqnarray} 
({M_\nu)}_{ij} \approx  \frac{f_3 f_5 v_1 v_2 u_1 u_4}{16 \pi^2} \sum_k {h_{N, \Sigma} }_{ik} {h_{N, \Sigma} }_{jk} \left( \frac{m_{2k}}{m^4_{sc}}\right)
\end{eqnarray}
Taking $u_1, u_4$ to be at 5 TeV and $m_{sc} \approx 500$ GeV, $v_1, v_2$ at electroweak scale and the singlet mass $m_{2k}$ at 100 GeV, the above expression can give rise to eV scale neutrino mass if 
$$ f_3 f_5 h_N h_N \sim 4 \times 10^{-11}$$
whereas for singlet mass $m_{2k}$ at 35 GeV, this constraint becomes 
\begin{equation}
f_3 f_5 h_N h_N \sim 1.12 \times 10^{-10}
\label{nuconstraint}
\end{equation}
One may note here that the appropriate explanation for the Gamma ray excess from the galactic center can be explained if one considers $m_{N_R} \approx m_{S2R} \approx m_{2k} = 35$ GeV \cite{Berlin:2014tja}. 

The expressions for neutrino masses and mixing angles can be derived using the parameters of the model which appear in the Lagrangian written for the model with additional $U(1)_X$ gauge symmetry. The value of these parameters change while going from high energy scale with unbroken $U(1)_X$ down to electroweak scale due to the effects of RGE. The corresponding changes in the neutrino parameters under RGE for the original Ma model were studied by \cite{merle2}. The authors have shown that the RGE effects on neutrino parameters can be quite large in these models, due to the dependence of neutrino parameters on products of several couplings of the model
when neutrinos obtain masses only via one-loop diagrams. However, in our case, there is tree level contribution also to the neutrino mass matrix and in renormalising the theory the counter-terms of the broken electroweak phase can be obtained from the symmetric phase by simple algebraic relations \cite{renorm1}. So the analysis done in \cite{merle2} is not directly applicable in our case.  However, a more complete analysis of neutrino masses and mixing in our model should include RGE effects.

\begin{figure}[h!] 
\centering 
\begin{tabular}{c}
\epsfig{file=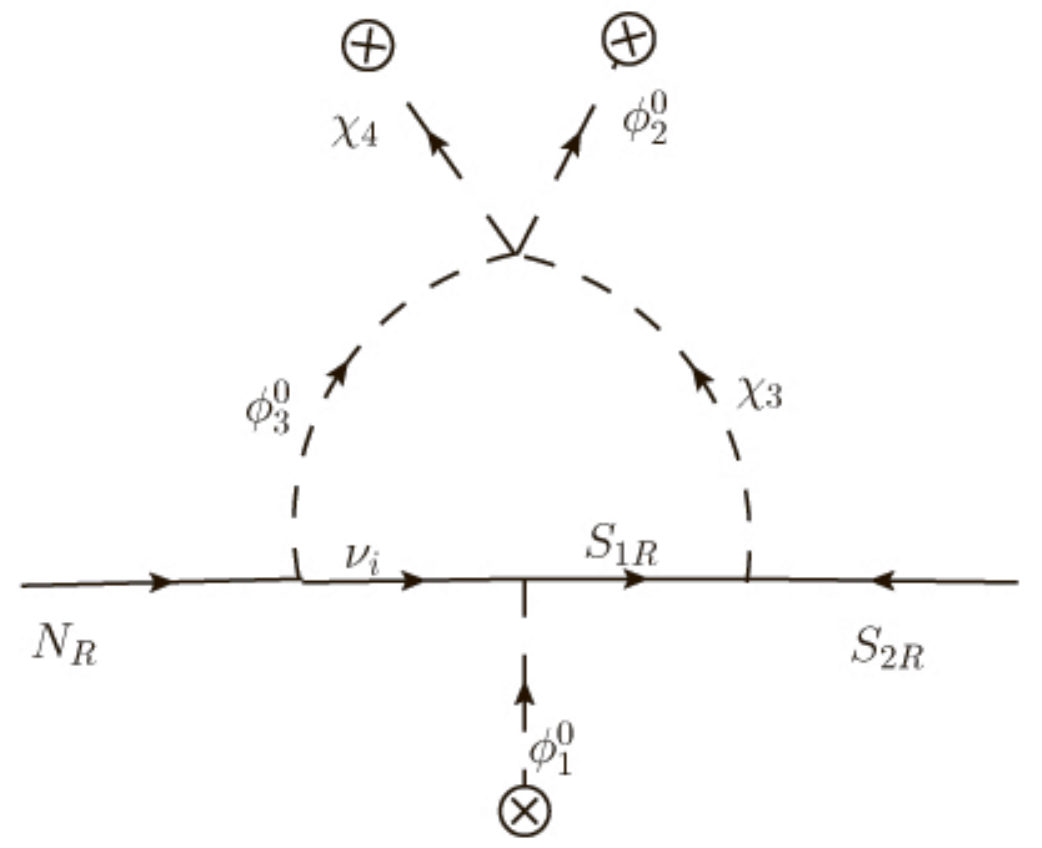,width=0.5\linewidth,clip=}\\
\end{tabular}
\caption{Mass splitting between $N_R$ and $S_{2R}$ at one-loop level.}
\label{fig9}
\end{figure}

\section{Masses of dark matter $N_R$ and $S_{2R}$}
\label{split}
For the discussion on dark matter, we assumed the mass of the heavier dark matter candidate $N_R$ to be 35 GeV, 
while keeping the lighter dark matter mass lower by about 3.55 keV. From the Yukawa Lagrangian of the model in equation \eqref{yukawa}, it can be seen that $N_R$ and $S_{2R}$ receive tree level masses from the vev's of $\chi_4$ and $\chi_2$ respectively. 
Considering Yukawa couplings $f_{S2}$ and $f_N$ of the order of $1.7\times10^{-2}$ and the vev's of $\chi_2$ and $\chi_4$ of the order of $2$ TeV,
with slight difference in the values of these Yukawa couplings, it is possible to generate mass differences of $3.55$ keV between $N_R$ and $S_{2R}$ at the tree level.

However such small mass difference at the tree level could change if there is corrections from higher order or loop effects. Here we have checked whether the higher order one-loop diagram as shown in figure \ref{fig9} could potentially contribute and change the tree level estimate 
of the mass difference between $S_{2R}$ and $N_R$. 
The mass matrix $M_{dark}$ in the $N_R, S_{2R}$ basis can be written as
\begin{equation}
M_{dark} =
\left(\begin{array}{cc}
\ f_N u_4 & {M_{dark}}_{12}  \\
\ {M_{dark}}_{21} & f_{S2} u_2 
\end{array}\right)
\label{numassmatrix}
\end{equation}
The  one-loop contribution ${M_{dark}}_{12} = {M_{dark}}_{21}^*$ to $2\times 2$  mass matrix is given by

\begin{eqnarray} 
{M_{dark}}_{12} && \approx   \frac{ 2 y^2 y^* h_N f_{12} {f_5}^* v_1^3 v_2  u_4}{16 \pi^2} \nonumber \\ && \left[ I\left( m_{\phi^0_{3R}},m_{\chi_{3R}}, m_\nu, m_{S_{1R}} \right) - I\left(m_{\phi^0_{3I}},m_{\chi_{3I}}, m_\nu, m_{S_{1R}} \right) \right] 
\label{darkmasssplit1}
\end{eqnarray}
in which
\begin{eqnarray}
I(a,a,b,c) \approx  \frac{a^2 \ln (a^2/c^2) -a^2 +c^2}{a^2 {(a^2 -c^2)}^2},
\label{darkmasssplit2}
\end{eqnarray}
for $b << a, c$  and 
\begin{eqnarray} 
I(a,b,c,d) \approx \frac{1}{a^2-b^2}  \left[  \frac{1}{a^2-d^2}  \ln(a^2/d^2) - \frac{1}{b^2-d^2} \ln(b^2/d^2)\right]
\label{darkmasssplit3}
\end{eqnarray}
for $c << a, b, d$ and these limits are useful as active neutrino mass scale $m_\nu$ is very small in comparison to other masses.  Here, $m_{\phi^0_{3R}}$ and $m_{\phi^0_{3I}} $ are the masses corresponding to $Re[\phi^0_3]$ and $Im[\phi^0_3]$ respectively whereas $m_{\chi_{3R}}$ and  $m_{\chi_{3I}}$ are the masses corresponding to $Re[\chi^0_3]$ and $Im[\chi^0_3]$ respectively. As denoted in the previous section, $v_i$ and $u_i$ are the vevs' of electroweak scalar doublets and additional singlet scalar fields respectively. We have considered $m_\nu \sim \frac{2 y^2 v_1^2}{f_S u_1}$
and $m_{S_{1R}} \sim f_s u_1$. 

Particularly, if we consider $ m_{\phi^0_{3}} \approx m_{\chi_{3}}$ and denote the near-equal masses of $Re[m_{s3}]$ and $Im[m_{s3}]$ as $m_s$ 
where $s_3 = \phi_3, \chi_3$, then  
\begin{eqnarray} 
{M_{dark}}_{12} && \approx   \frac{ 2 y^2 y^* h_N f_{12} {f_5}^* v_1^3 v_2  u_4}{16 \pi^2} \nonumber \\ && \left[ 
\frac{\left( m^2_{\phi^0_{3R}} - m^2_{\phi^0_{3I}} \right) \{ \ln \left( m_s^2/(f_s^2 u_1^2)\right ) -1  \} }{m_s^2 {(m_s^2- f_s^2 u_1^2)}^2} \right] 
\label{darkmasssplit4}
\end{eqnarray}
From tree level neutrino mass $m_\nu \sim \frac{2 y^2 v_1^2}{f_S u_1}$, if we keep the mass of singlet neutrino $m_{S_{1R}} \sim f_s u_1$ 
fixed at 2 TeV or so as discussed just after eq. \eqref{neutmass}, then the Yukawa couplings have to be of the order of $3 \times 10^{-6}$ to give rise to neutrino mass of order $0.1$ eV. 
Considering $m_s =(m_{\phi^0_{3R}} + m_{\phi^0_{3I}})/2 \approx 500$ GeV, $m_{\phi^0_{3R}} - m_{\phi^0_{3I}} \approx 10$ GeV and other parameters 
mentioned above we obtain   
\begin{eqnarray} 
{M_{dark}}_{12} && \approx  f_5f_{12}h_N \times 10^{-16} \;{\rm GeV}
\label{darkmasssplit5}
\end{eqnarray}
Even if we consider the couplings $f_5$,$f_{12}$ and $h_N$ of $\mathcal{O}(1)$ then also ${M_{dark}}_{12} \approx 10^{-16}$ GeV which 
is $\mathcal{O}(10^{10})$ less than the mass splitting of $3.55$ keV as considered at the tree level. 
Therefore, with our choices of parameters, ${M_{dark}}_{12}$ is negligible in \eqref{numassmatrix} and hence the higher order one-loop correction does not alter the tree level mass difference of $N_R$ and $S_{2R}$.

To check the compatibility of different requirments associated with
\begin{enumerate}
\item the condition on the product of couplings from the decay of the heavier dark matter producing the $3.55$ keV line. 
\item obtaining appropriate heavier neutrino mass from the tree level.
\item obtaining appropriate lighter neutrino mass from tree level as well as one-loop level. 
\item keV mass difference of the two dark matter $N_R$ and $S_{2R}$ at the tree level. 
\end{enumerate}
we analyse the conditions in equations \eqref{xrayconstraint}, \eqref{neutmass}, \eqref{nuconstraint} and \eqref{darkmasssplit5} respectively.
Although the equation \eqref{darkmasssplit5} is trivially satisfied with any value of Yukawa couplings less than $\mathcal{O}(1)$ but other 
equations give certain conditions on the Yukawa couplings as well as conditions of $CP$ eigenvalues of the two dark matter particles. From equation \eqref{neutmass}, 
with our appropriate choice of parameters we have obtained $y\sim 3\times 10^{-6}$. Considering this value of $y$ we find that the condition given in 
equation \eqref{xrayconstraint} can be satisfied only for the opposite $CP$ eigenvalues of two dark matter particles as other Yukawa couplings in that equation 
are expected to be of $\mathcal{O}(1)$ or less. Considering the condition of opposite $CP$ eigenvalues in equation \eqref{xrayconstraint} and 
using the value of $y$ as obtained from equation \eqref{neutmass} we get $h_N \sim \frac{10^{-8}}{f_6f_{12}}$
 using which in equation \eqref{nuconstraint} we obtain the following condition on the ratio of different Yukawa couplings
\begin{align}
\frac{f^2_6f^2_{12}}{f_3f_5} \sim 10^{-6}
\end{align}
Considering Yukawa couplings like $f_6,f_{12},f_5$ and $f_3$ of the same order they turn out to be $\mathcal{O}(10^{-3})$ whereas 
$h_N\sim \mathcal{O}(10^{-2})$.
So, in our model with suitable values of the Yukawa couplings as discussed above, it is possible to explain the 355 keV X-ray line, tiny neutrino mass and gamma ray excess from galactic centre in a single framework.

\section{Results and Conclusion}
\label{conclude}
We have studied an abelian gauge extension of the standard model which can predict tiny neutrino mass and stable dark matter candidate naturally. In particular, we have discussed the possibility of explaining the recently observed 3.55 keV X-ray line and the galactic center gamma ray excess from a common dark matter origin within the framework of this abelian gauge model. Although the model has both scalar and fermionic dark matter candidates, we choose to study only fermionic dark matter candidates to serve our goal better. The dark matter candidate is guaranteed to be stable by a remnant $Z_2$ symmetry after the abelian gauge symmetry gets spontaneously broken. In order to explain the 3.55 keV X-ray line, we assume the dark sector to consist of two dark matter particles: the lightest $Z_2$-odd particle $(S_{2R})$ and the next-to-lightest $Z_2$-odd particle $N_R$, both of which are singlet fermions. The mass difference between the two dark matter particles is chosen to be 3.55 keV such that the heavier one can decay into the lighter one and a photon at loop level. In order to explain the galactic center gamma ray excess, we choose the lightest dark matter mass to be 35 GeV and check whether the two dark matter particles give rise to the required annihilation cross sections. We also take into account the constraints from dark matter direct detection experiments like XENON100 on spin dependent scattering cross section of dark matter off nuclei. These models can also face stringent limits on new gauge boson mass $M_X$ and gauge coupling $g_X$. Using the results from \cite{collider} where the authors found the lower bound on $M_X/g_X$ to be $2.6$ TeV for $\text{BR}(X \rightarrow \text{SM}) = 90\%$ and $g_X = 0.1$ we also use three different cuts on $M_X/g_X$ starting from a moderate 2 TeV to a conservative 4 TeV on $M_X/g_X$. These limits will be even weaker in those region of parameter space where $g_X$ can be much lower than $0.1$. We find that, even after applying a conservative lower limit on $M_X/g_X$ as $4$ TeV, we still have some parameter space available near the s-wave resonance region which can satisfy all constraints related to dark matter and colliders.
 
After showing the allowed parameter space in terms of $g_X, M_X$ as well as $u$, the common vev of the scalar singlets, we constrain the other 
parameters of the model from the requirement of producing the correct 3.55 keV X-ray flux, sub-eV neutrino mass and keeping one-loop mass splitting 
between two dark matter candidates below keV. From the X-ray flux constraints, we find that the product of four relevant dimensionless couplings have to be 
around $1$ or $10^{-14}$ for same $CP$ eigenvalues or opposite $CP$ eigenvalues respectively for $N_R$ and $S_{2R}$. Similarly, the constraints 
from sub-eV neutrino masses keep the product of four relevant dimensionless couplings tuned at around $10^{-10}$ for singlet fermion dark matter masses of a few tens of GeV. 
 Constraints from tree level light neutrino mass and decay of heavier dark matter to 
 lighter dark matter and photon show that only opposite $CP$ 
eigenvalues of $N_R$ and $S_{2R}$ could be possible for Yukawa couplings of $\mathcal{O}(1)$ or less. 
In our model the small mass difference between two dark matters of $3.55$ keV considered at the tree level remains unchanged even with higher 
order corrections.
Also the allowed parameter space in $g_X-M_X$ plane is very limited. This is because for light dark matter mass, in order to explain GC excess and 3.55 keV X-Ray line together, the constraints from dark matter experiments as well as bound on $M_X/g_X$ allow only a limited region near the s-channel resonance $M_X \approx \; 2 m_{DM}$. For dark matter mass around 35 GeV, the allowed mass of neutral boson becomes around 70 GeV which again faces severe constraints from LEP-II data and further constrain the coupling $g_X \leq 10^{-2}$. Due to the very limited parameter space available, this model will undergo serious scrutiny at future experiments with more sensitivity.

\begin{acknowledgments}
DB would like to thank the organizers of the workshop "LHCDM-2015" during 9-28 February, 2015 at IACS, Kolkata, India where some important discussions related to this work took place. AD likes to thank Council of Scientific and Industrial Research, Government of India for financial support through Senior Research Fellowship.
\end{acknowledgments}

\bibliographystyle{apsrev}

\begin{thebibliography}{30}
\expandafter\ifx\csname natexlab\endcsname\relax\def\natexlab#1{#1}\fi
\expandafter\ifx\csname bibnamefont\endcsname\relax
  \def\bibnamefont#1{#1}\fi
\expandafter\ifx\csname bibfnamefont\endcsname\relax
  \def\bibfnamefont#1{#1}\fi
\expandafter\ifx\csname citenamefont\endcsname\relax
  \def\citenamefont#1{#1}\fi
\expandafter\ifx\csname url\endcsname\relax
  \def\url#1{\texttt{#1}}\fi
\expandafter\ifx\csname urlprefix\endcsname\relax\def\urlprefix{URL }\fi
\providecommand{\bibinfo}[2]{#2}
\providecommand{\eprint}[2][]{\url{#2}}

\bibitem{Xray1} E.~Bulbul,M.~Markevitch,A.~Foster,R.~K.~Smith,M.~Loeewenstein and S.~W.~Randall,
Astrophys.\ J.\ {\bf 789}, 13 (2014)
[arXiv:1402.2301 [astro-ph.CO]].
\bibitem{Xray2} A. Boyarsky, O. Ruchayskiy, D. Iakubovskyi and J. Franse, Phys. Rev. Lett. {\bf 113}, 251301 (2014).
\bibitem[{\citenamefont{Jungman et~al.}(1996)\citenamefont{Jungman,
  Kamionkowski, and Griest}}]{Jungman:1995df}
\bibinfo{author}{\bibfnamefont{G.}~\bibnamefont{Jungman}},
  \bibinfo{author}{\bibfnamefont{M.}~\bibnamefont{Kamionkowski}},
  \bibnamefont{and} \bibinfo{author}{\bibfnamefont{K.}~\bibnamefont{Griest}},
  \bibinfo{journal}{Phys. Rept.} \textbf{\bibinfo{volume}{267}},
  \bibinfo{pages}{195} (\bibinfo{year}{1996}), \eprint{hep-ph/9506380}.
\bibitem{Xraysterile} H. Ishida, K. S. Jeong and F. Takahashi, Phys. Lett. {\bf B732}, 196 (2014); K. N. Abazajian, Phys. Rev. Lett. {\bf 112}, 161303 (2014); S. Baek and H. Okada, arXiv:1403.1710; B. Shuve and I. Yavin, Phys. Rev. {\bf D89}, 113004 (2014); T. Tsuyuki, Phys. Rev. {\bf D90}, 013007 (2014); F. Bezrukov and D. Gorbunov, Phys. Lett. {\bf B736}, 494 (2014); D. J. Robinson and Y. Tsai, Phys. Rev. {\bf D90}, 045030 (2014); S. Chakraborty, D. K. Ghosh and S. Roy, JHEP {\bf 1410}, 146 (2014); N. Haba, H. Ishida and R. Takahashi, arXiv:1407.6827; S. Patra and P. Pritimita, arXiv:1409.3656; A. Merle and A. Schneider, arXiv:1409.6311; S. K. Kang and A. Patra, arXiv:1412.4899.
\bibitem{Xrayothers1} D. P. Finkbeiner and N. Weiner, arXiv:1402.6671; T. Higaki, K. S. Jeong and F. Takahashi, Phys. Lett. {\bf B733}, 25 (2014); J. Jaeckel, J. Redondo and A. Ringwald, Phys. Rev. {\bf D89}, 103511 (2014); H. M. Lee, S. C. Park and Wan-II Park, Eur. Phys. J. {\bf C74}, 3062 (2014); R. Krall, M. Reece and T. Roxlo, JCAP {\bf 1409}, 007 (2014); J.-C. Park, S. C. Park and K. Kong, Phys. Lett. {\bf B733}, 217 (2014); M. T. Frandsen, F. Sannino and O. Svendsen, JCAP {\bf 1405}, 033 (2014); K. Nakayama, F. Takahashi and T. Yanagida, Phys. Lett. {\bf B735}, 338 (2014); K.-Y. Choi and O. Seto, Phys. Lett. {\bf B735}, 92 (2014); M. Cicoli, J. P. Conlon, M. C. D. Marsh and M. Rummel, Phys. Rev. {\bf D90}, 023540 (2014); C. Kolda and J. Unwin, Phys. Rev. {\bf D90}, 023535 (2014); R. Allahverdi, B. Dutta and Y. Gao, Phys. Rev. {\bf D89}, 127305 (2014); N. -E. Bomark and L. Roszkowski, Phys. Rev. {\bf D90}, 011701 (2014); S. P. Liew, JCAP {\bf 1405}, 044 (2014); K. Nakayama, F. Takahashi and T. T. Yanagida, Phys. Lett. {\bf B734}, 178 (2014).
\bibitem{Xrayothers2} F. S. Queiroz and K. Sinha, Phys. Lett. {\bf B735}, 69 (2014); E. Dudas, L. Heurtier and Y. Mambrini, Phys. Rev. {\bf D90}, 035002 (2014); K. S. Babu and R. N. Mohapatra, Phys. Rev. {\bf D89}, 115011 (2014); K. P. Modak, arXiv:1404.3676; J. M. Cline, Y. Farzan, Z. Liu, G. D. Moore and W. Xue, Phys. Rev. {\bf D89}, 121302 (2014); H. Okada and T. Toma, Phys. Lett. {\bf B737}, 162 (2014); J. P. Conlon and F. V. Day, JCAP {\bf 11}, 033 (2014); S. Baek, P. Ko and Wan-II Park, arXiv:1405.3730; N. Chen, Z. Liu and P. Nath, Phys. Rev. {\bf D90}, 035009 (2014); J. P. Conlon and A. J. Powell, arXiv:1406.5518; H. Ishida and H. Okada, arXiv:1406.5808; C. -Q. Geng, D. Huang and L. -H. Tsai, JHEP {\bf 1408}, 086 (2014); B. Dutta, I. Gogoladze, R. Khalid and Q. Shafi, JHEP {\bf 1411}, 018 (2014); H. Okada and Y. Orikasa, Phys. Rev. {\bf D90}, 075023 (2014); J. M. Cline and A. R. Frey, JCAP {\bf 1410}, 013 (2014); Y. Farzan and A. R. Akbarieh, JCAP {\bf 1411}, 015 (2014); K. K. Boddy, J. L. Feng, M. Kaplinghat, Y. Shadmi and T. M. P. Tait, Phys. Rev. {\bf D90}, 095016 (2014); K. Schutz and T. R. Slatyer, arXiv:1409.2867; J. M. Cline and A. R. Frey, arXiv:1410.7766; K. Cheung, W. -C. Huang and Y. -L. S. Tsai, arXiv:1411.2619; A. Harada, A. Kamada and N. Yoshida, arXiv:1412.1592; G. Arcadi, L. Covi and F. Dradi, arXiv:1412.6351; A. Biswas, D. Majumdar and P. Roy, arXiv:1501.02666; A. Berlin, A. DiFranzo and D. Hooper, arXiv:1501.03496.
\bibitem{Xrayweakscale} Z. Kang, P. Ko, T. Li and Y. Liu, arXiv:1403.7742; H. M. Lee, Phys. Lett. {\bf B738}, 118 (2014); C. -W. Chiang and T. Yamada, JHEP {\bf 1409}, 006 (2014);  S. Baek, arXiv:1410.1992; S. Patra, N. Sahoo and N. Sahu, arXiv:1412.4253; H. M. Lee, C. B. Park and M. Park, arXiv:1501.05479.
\bibitem{Falkowski:2014sma} A.~Falkowski,Y.~Hochberg and J.~T.~Ruderman,
JHEP {\bf 1411}, 140 (2014)[arXiv:1409.2872 [hep-ph]].
\bibitem{Planck13} P.~A.~R.~Ade \textit{et al.} [Planck Collaboration], Astron. Astrophys. {\bf 571}, A16 (2014).

\bibitem[{\citenamefont{Adhikari et~al.}(2009)\citenamefont{Adhikari, Erler,
  and Ma}}]{Adhikari:2008uc}
\bibinfo{author}{\bibfnamefont{R.}~\bibnamefont{Adhikari}},
  \bibinfo{author}{\bibfnamefont{J.}~\bibnamefont{Erler}}, \bibnamefont{and}
  \bibinfo{author}{\bibfnamefont{E.}~\bibnamefont{Ma}}, \bibinfo{journal}{Phys.
  Lett.} \textbf{\bibinfo{volume}{B672}}, \bibinfo{pages}{136}
  (\bibinfo{year}{2009}), \eprint{0810.5547}.
  
  \bibitem{Borah:2012qr}
D.~Borah and R.~Adhikari, {Phys. Rev.} {\bf D85}, 095002 (2012), 1202.2718.

\bibitem{Borah:2014} D. Borah and R. Adhikari, Phys. Lett. {\bf B729}, 143 (2014); R. Adhikari, D. Borah and E. Ma, arXiv:1411.4602.
  
\bibitem[{\citenamefont{Ma}(2006)}]{Ma:2006km}
\bibinfo{author}{\bibfnamefont{E.}~\bibnamefont{Ma}}, \bibinfo{journal}{Phys.
  Rev.} \textbf{\bibinfo{volume}{D73}}, \bibinfo{pages}{077301}
  (\bibinfo{year}{2006}), \eprint{hep-ph/0601225}.
  \bibitem{arnabborah2} D. Borah and A. Dasgupta, Phys. Lett. {\bf B741}, 103 (2015).
  \bibitem{GCfeb} T. Daylan, D. P. Finkbeiner, D. Hooper, T. Linden, S. K. N. Portillo, N. L. Rodd and T. R. Slatyer, arXiv:1402.6703.
\bibitem{arnabborah} A. Dasgupta and D. Borah, Nucl. Phys. {\bf B889}, 637 (2014).

\bibitem[{\citenamefont{Ma and Suematsu}(2009)}]{Ma:2008cu}
\bibinfo{author}{\bibfnamefont{E.}~\bibnamefont{Ma}} \bibnamefont{and}
  \bibinfo{author}{\bibfnamefont{D.}~\bibnamefont{Suematsu}},
  \bibinfo{journal}{Mod.Phys.Lett.} \textbf{\bibinfo{volume}{A24}},
  \bibinfo{pages}{583} (\bibinfo{year}{2009}), \eprint{0809.0942}.



 \bibitem[{\citenamefont{Kolb and Turner}(1990)}]{Kolb:1990vq}
\bibinfo{author}{\bibfnamefont{E.~W.} \bibnamefont{Kolb}} \bibnamefont{and}
  \bibinfo{author}{\bibfnamefont{M.~S.} \bibnamefont{Turner}},
  \bibinfo{journal}{Front. Phys.} \textbf{\bibinfo{volume}{69}},
  \bibinfo{pages}{1} (\bibinfo{year}{1990}).


\bibitem[{\citenamefont{Gondolo and Gelmini}(1991)}]{Gondolo:1990dk}
\bibinfo{author}{\bibfnamefont{P.}~\bibnamefont{Gondolo}} \bibnamefont{and}
  \bibinfo{author}{\bibfnamefont{G.}~\bibnamefont{Gelmini}},
  \bibinfo{journal}{Nucl. Phys.} \textbf{\bibinfo{volume}{B360}},
  \bibinfo{pages}{145} (\bibinfo{year}{1991}).
\bibitem{Berlin:2014tja} A. Berlin, D. Hooper and S. D. McDermott, Phys. Rev. {\bf D89}, 115022 (2014).
\bibitem{Aprile:2013doa} E. Aprile et al. Phys. Rev. Lett. {\bf 111} (2), 021301 (2013).  
\bibitem{pdg} J. Beringer \textit{et al.}, Phys. Rev. {\bf D86}, 010001 (2012).
\bibitem{collider} G. Arcadi, Y. Mambrini, M. H. G. Tytgat and B. Zaldivar, JHEP {\bf 1403}, 134 (2014).
\bibitem{Kayser:1984ge} 
  B.~Kayser,
  Phys.\ Rev.\ D {\bf 30}, 1023 (1984).
  
\bibitem{Pal:1981rm} 
  P.~B.~Pal and L.~Wolfenstein,
  Phys.\ Rev.\ D {\bf 25}, 766 (1982).

\bibitem{GCprev} L. Goodenough and D. Hooper, arXiv:0910.2998; A. Boyarsky, D. Malyshev and O. Ruchayskiy, Phys. Lett. {\bf B705}, 165 (2011); D. Hooper and T. Linden, Phys. Rev. {\bf D84}, 123005 (2011); K. N. Abazajian and M. Kaplinghat, Phys. Rev. {\bf D86}, 083511 (2012); C. Gordon and O. Macias, Phys. Rev. {\bf D88}, 083521 (2013); D. Hooper and T. R. Slatyer, Phys. Dark. Univ. {\bf 2}, 118 (2013); K. N. Abazajian, N. Canac, S. Horiuchi and M. Kaplinghat, Phys. Rev. {\bf D90}, 023526 (2014).
\bibitem{merle1} A. Merle and P. Platscher, arXiv:1502.03098.
\bibitem{ti}
P.~Minkowski,
{Phys. Lett.} {\bf B67}, 421 (1977);
M.~Gell-Mann, P.~Ramond, and R.~Slansky (1980), print-80-0576 (CERN);
T.~Yanagida (1979), in Proceedings of the Workshop on the Baryon Number of the Universe and Unified Theories, Tsukuba, Japan, 13-14 Feb 1979;
R. N.~Mohapatra and G.~Senjanovic,
{Phys. Rev. Lett} {\bf 44}, 912 (1980);
J.~Schechter and J. W. F.~Valle,
{Phys. Rev.} {\bf D22}, 2227 (1980).
\bibitem{merle2} R. Bouchand and A. Merle, JHEP {\bf 1207}, 084 (2012); A. Merle and M. Platscher, arXiv:1507.06314.
\bibitem{renorm1} J. C. Collins,
Renormalization:  An Introduction to Renormalization, the Renormal-
ization  Group  and  the  Operator-Product  Expansion
(Cambridge  University  Press,
1984), ISBN 9780521311779.
\end{thebibliography}

\end{document}